\documentclass[11pt]{article}

\usepackage[hyperfootnotes=false]{hyperref}
\usepackage{fullpage}

\hypersetup{
    pdftitle={Matrix Element Method in HEP: Transfer Functions, Efficiencies,
              and Likelihood Normalization},
    pdfauthor={Igor Volobouev},
    pdfsubject={Data Analysis, Statistics and Probability},
    urlbordercolor={0 0 1},
    bookmarksnumbered=true
}

\newcommand{\sub}[1]{\ensuremath{_{\mbox{\scriptsize \,#1}}}}

\newenvironment{thinlist} {
    \begin{list} {---} {
        \setlength{\topsep}{0.075cm}
        \setlength{\parsep}{0.075cm}
        \setlength{\itemsep}{0.075cm}
    }
} {\end{list}}

\begin{document}
\title{Matrix Element Method in HEP: Transfer Functions, Efficiencies,
and Likelihood Normalization}

\author{Igor~Volobouev \vspace{0.3cm} \\
{\it i.volobouev@ttu.edu} \vspace{0.3cm} \\
Texas Tech University, Department of Physics, Box 41051,\\
Lubbock, Texas, USA 79409
}

\maketitle

\begin{abstract}
This article surveys the procedures used for deriving detector
transfer functions
and normalizing probability densities for the statistical analysis technique
known as the ``matrix element method'' in the context of
high energy physics (HEP) data analysis. Common misconceptions
about transfer functions and efficiencies are pointed out and clarified.
\end{abstract}

\section{Matrix Element Analysis in a Nutshell}
\label{sec:form1}

The matrix element (ME) method was introduced into
the HEP data analysis practice relatively recently, 
in a pioneering measurement of the top quark mass by the
D0~Collaboration~\cite{ref:estrada_thesis,ref:d0}.
The method can be utilized both in  searches of new phenomena
and in precision measurements of physical model parameters.
It consists in calculating
the probability of observing an event in a particle detector
according to
\begin{eqnarray}
P\sub{ev}({\bf y} | {\bf a}) = \sum_{i} f_{i} P_{i}({\bf y} | {\bf a})
\end{eqnarray}
where ${\bf y}$ is the vector of observed quantities
in the measurement space $Y\!$,
${\bf a}$~is the vector of model parameters (both theoretical
and instrumental), and $f_{i}$ are fractions of different
non-interfering production channels consistent with ${\bf y}$.
In the following discussion, it will be assumed that
$f_{i} \ge 0$ for every production channel $i$ and that
the constraint $\sum_{i} f_{i} = 1$ is imposed.
For each channel $i$, the probability to measure ${\bf y}$
is estimated from
\begin{eqnarray}
{
P_{i}({\bf y} | {\bf a}) = \frac{\Omega({\bf y})}{\sigma_{i}({\bf a}) A_{i}({\bf a})}
\int_{X_i} W_{i}({\bf y} | {\bf x}, {\bf a})\,
\epsilon_{i} ({\bf x}, {\bf a})\,
|M_{i} ({\bf x}, {\bf a})|^{2} \,T_{i}({\bf x}, {\bf a}) \,d {\bf x}
\label{eq:prob_integral}
}
\end{eqnarray}
where

$\Omega({\bf y})$ --- Indicator function
for the analysis acceptance (1 for events which pass the
event selection criteria, 0 otherwise). This term can be
replaced by 1 in case only accepted events are considered.

${\bf x}$ --- Variables which uniquely specify a point in
              the channel phase space $X_i$, ${\bf x} \in X_i$.

$d {\bf x}$ --- Differential element of the phase space $X_i$.

$\sigma_{i}({\bf a})$ --- Channel cross section: $\sigma_{i}({\bf a}) = \int_{X_i} 
|M_{i} ({\bf x}, {\bf a})|^{2} \,T_{i}({\bf x}, {\bf a}) \,d {\bf x}$.

$A_{i}({\bf a})$ --- Overall experimental acceptance.

$W_{i}({\bf y} | {\bf x}, {\bf a})$ --- Detector transfer function.
This is the probability density to observe detector response ${\bf y} \in Y$ when
the ``true'' phase space coordinate of the event is ${\bf x}$.
This function is normalized by
$\int_{Y} \Omega({\bf y}) W_{i}({\bf y} | {\bf x}, {\bf a})\, d {\bf y} = 1$
for every value of $i$, ${\bf x}$, and ${\bf a}$.

$\epsilon_{i} ({\bf x}, {\bf a})$ --- Efficiency to detect an event
originated at the phase space point~${\bf x}$ ({\it i.e.,} the probability
that an event originated at ${\bf x}$ actually ends up with ${\bf y}$
for which $\Omega({\bf y}) = 1$).

$|M_{i} ({\bf x}, {\bf a})|^{2}$ --- Squared matrix element of the process.

$T_{i}({\bf x}, {\bf a})$ --- Other factors which do not depend 
on ${\bf y}$
({\it e.g.,} flux of colliding beams, parton distribution functions).

The normalization condition imposed on $W_{i}({\bf y} | {\bf x}, {\bf a})$
ensures that $P_{i}({\bf y} | {\bf a})$ is a properly normalized
probability density in $Y$ for all values of ${\bf a}$. Indeed,
$$
\int_{Y} P_{i}({\bf y} | {\bf a}) \,d {\bf y} = \int_{Y} \frac{\Omega({\bf y})}{\sigma_{i}({\bf a}) A_{i}({\bf a})}
\int_{X_i} W_{i}({\bf y} | {\bf x}, {\bf a})\,
\epsilon_{i} ({\bf x}, {\bf a})\,
|M_{i} ({\bf x}, {\bf a})|^{2} \,T_{i}({\bf x}, {\bf a}) \,d {\bf x} \,d {\bf y}.
$$
Integrating over ${\bf y}$ first and taking into account
the normalization of the transfer function, one gets
$$
\int_{Y} P_{i}({\bf y} | {\bf a}) \,d {\bf y} = \frac{1}{\sigma_{i}({\bf a}) A_{i}({\bf a})}
\int_{X_i} \epsilon_{i} ({\bf x}, {\bf a})\,
|M_{i} ({\bf x}, {\bf a})|^{2} \,T_{i}({\bf x}, {\bf a}) \,d {\bf x} = 
\frac{\langle \epsilon_{i} ({\bf x}, {\bf a}) \rangle}{A_{i}({\bf a})} \equiv 1,
$$
where the symbol $\langle ... \rangle$ stands for averaging with
respect to the process density $p({\bf x} | {\bf a})$ in the phase space:
$p({\bf x} | {\bf a}) = \frac{1}{\sigma_{i}({\bf a})} |M_{i} ({\bf x}, {\bf a})|^{2} \,T_{i}({\bf x}, {\bf a})$. With properly normalized $P_{i}({\bf y} | {\bf a})$, normalization of
$P\sub{ev}({\bf y} | {\bf a})$ is ensured as well.

The formulation of the ME method just described
assumes that the production channel fractions
$f_{i}$ are known with sufficient precision and are
not of interest in the data analysis.
It can be easily
generalized for use with some densities $\pi (f_{i}|{\bf a})$ which
represent the prior knowledge about $f_{i}$ as well as to the case
in which $f_{i}$ are parameters to be measured.

The ME approach offers several important advantages
over all other data analysis schemes commonly used in HEP:
\begin{thinlist}
\item The method is universal and can be applied to a wide variety
      of particle processes for which theoretical models have been established.
\item The theoretical assumptions about the process under study
      (matrix elements $M_{i} ({\bf x}, {\bf a})$, channel
      fractions $f_{i}$, parton distribution functions)
      are incorporated into the data analysis in
      the most efficient manner. The whole procedure can be 
      viewed as a~Bayesian marginalization of the event probability over all unobserved
      degrees of freedom. Particle theory provides a~well-motivated informative prior
      for this marginalization.
\item Some widely used data analysis methods introduce
      implicit assumptions about the shape of detector resolution functions.
      In particular, methods based on $\chi^2$ minimization (used, {\it e.g.}, in kinematic fitting)
      assume
      Gaussian measurement errors, and the effect of this assumption on the quality
      of statistical modeling can not be quantified within the $\chi^2$-based method itself.
      There is no such inherent restriction in the ME approach; very detailed and precise
      detector models can be usefully employed.
\item Maximization of the likelihood 
      $L({\bf a}) = \prod P\sub{ev}({\bf y} | {\bf a})$
      results in an efficient (in the statistical sense)
      estimate of the parameter {\bf a}. Straightforward
      profiling or marginalization
      of the likelihood can be
      utilized in case some of the ${\bf a}$ dimensions are not of interest
      and can be treated as nuisance parameters. In effect, systematic uncertainties
      (which are nothing else but the uncertainties due to imprecisely known values of
       nuisance parameters) are calculated {\it on the event-by-event basis} and,
       therefore, each event contributes into the overall parameter estimate
       with an optimal weight which takes into account both statistical and systematic uncertainty.
\item According to the Neyman-Pearson lemma, the likelihood ratio
      $$r_{j}({\bf y} | {\bf a}) = \frac{P_{j}({\bf y} | {\bf a})}{\sum_{i \ne j} f_{i} P_{i}({\bf y} | {\bf a})}$$
      is the optimal discriminant function for channel $j$. In particular,
      when channel $j$ represents the signal under study,
      this function can be used to assign events to either ``signal''
      or ``background'' category by testing whether
      $r_{j}({\bf y} | {\bf a}) >$~cutoff. This method
      achieves the best misclassification
      rate among all possible
      classifiers with the same signal selection efficiency.
\end{thinlist}
Despite all these advantages, due to its complexity and lack
of standard computational tools the ME approach is not 
necessarily the obvious first
choice among various HEP data analysis techniques.
Consider, for example, the most general case in which the integration region
$X_i$ in Eq.~\ref{eq:prob_integral} corresponds to the
relativistic phase space of all initial and final state particles,
and the space of measurements $Y\!$ corresponds to the complete data record produced
by an~experimental apparatus in a~triggered high energy
event. This general case will remain intractable for any foreseeable
future because the dimensionality of the integral and the complexity
of the integrand are just too high: thousands of particles can be
produced in a~high energy collision and then traced in a~detector with
millions of data acquisition channels. Reducing the dimensionality
of $X_i$ and $Y\!$ spaces to a manageable level while preserving
the information about parameters of interest is one of the most critical
issues in practical applications of the ME method.

\section{A Comment on Dimensionality Reduction}

The first and the most important dimensionality reduction stage
employed by all collider experiments is the process of
``event reconstruction'',
and ``jet reconstruction'' in particular. In all HEP
applications of the ME method developed so far,
the final state ``soft QCD''
processes (parton showering and hadronization) were combined
together with the detector response,
and this combination was subsequently modeled
empirically by the jet transfer functions.
This particular approach is well motivated
(at least at high c.m.s. energies such as those of the Tevatron and the LHC)
by the QCD factorization theorem and results in a drastic dimensionality
reduction of $X_i$ which then becomes the phase space of
parton-level quantities.
In addition to jets, the event reconstruction
procedure also produces other high-level ``physics objects'' (lepton
candidates, decay vertices, missing transverse energy, {\it etc.}), so
the measurement space $Y\!$ is typically associated with the variables
which describe such objects.

After this first and necessary dimensionality reduction
stage, calculation of the Eq.~\ref{eq:prob_integral} integral
can still remain a~formidable problem.
 For example, the
leading order parton phase space of the reaction
$p\bar{p} \rightarrow t\bar{t} \rightarrow W^{+}bW^{-}\bar{b} 
\rightarrow \ell \nu + 4$ jets, $\ell = e \mbox{\ or\ } \mu$ 
(this is the ``golden channel'' for measuring the top quark mass
at the Tevatron)
is 24-dimensional\footnote{Two initial and six final
state particles are described by 32 variables, but four energy-momentum conservation
laws and four well-known masses (of the initial partons, $\ell$ and $\nu$) result
in a trivial reduction to 24.},
while the space of reconstructed quantities is described by $\sim$20
variables\footnote{Three momentum components for $\ell$ and each
jet, and transverse distance of the secondary vertex to the beam axis for each jet.
$\nu$ escapes undetected. Two additional variables
should be included in the count
if the missing transverse energy is used for event selection.}.
Moreover, the integrand has a complex structure because of
the resonant nature of $t$ and $W$, and its efficient evaluation
requires a highly nontrivial phase space sampling scheme. Inclusion
of the processes beyond the leading order increases the
phase space integral complexity even further. Since the number of integrals
which has to be evaluated is usually quite large,\footnote{For example,
a recent measurement of the top quark mass~\cite{ref:plujan}
reports using a sample of 1,087 $t\bar{t}$ candidate events,
while the likelihood is calculated on a $32 \times 26$ 
rectangular grid of parameter values.
The number of Monte Carlo events which has to be
processed in order to calibrate such a measurement
is typically couple orders of magnitude higher.} additional dimensionality reduction
assumptions may be necessary for purely practical reasons: to simplify
the reaction kinematics and to speed up the convergence of the integration procedures.
The following ideas have been explored, in various combinations:
\begin{thinlist}
\item Neglect some of the production channels (especially if their corresponding 
      event fractions $f_i$ are expected to be small).
\item Assume that some or all of the partons are on shell.
\item Use tree-level, leading order matrix element and phase space.
\item Further simplify the matrix element. Such simplifications
      might use narrow width approximations for the resonances,
      ignore spin correlations in pair production, consider only valence
      quarks in the production mechanisms, {\it etc}.
\item Assume that initial partons have zero transverse momentum (hadron beams implied).
\item Assume that some event variables can be perfectly measured
      in the detector ({\it i.e.,} transfer functions for these variables
      are represented by Dirac delta functions).
\end{thinlist}
Each of these assumptions reduces the quality of the process statistical model
and results in either degraded precision of a parameter estimator
(in case the purpose of the analysis is a parameter measurement) or diminished
power of a signal discriminant. Naturally, assumptions of this kind should
be avoided unless either their effect could be shown to be negligible or
the practical problems could not be overcome by other means.
A number of techniques could be employed to increase the calculation speed without
sacrificing the fidelity of the statistical model. These techniques include
optimization of phase space and channel
sampling~\cite{ref:madweight,ref:klepitau}, use of multidimensional integration
algorithms with better convergence properties than the standard
Monte Carlo~\cite{ref:Niederreiter,ref:qmcconv3,ref:Severino,ref:hiquasi},
and speeding up the evaluation of the integrand ({\it e.g.}, by
tabulating detector transfer functions for fast lookup, or by
factorizing the integrand into terms which depend on different parameters so that
some terms do not have to be recalculated for every value of ${\bf a}$).

\section{Alternative Formulations of the Method}
\label{sec:form2}

The first~\cite{ref:estrada_thesis,ref:d0}
and some subsequent ({\it e.g.,}~\cite{ref:canelli_thesis})
applications of the ME approach
in HEP data analysis utilized a~somewhat different formulation
of the method. Expressed with the notation introduced in Section~\ref{sec:form1},
the ``extended likelihood'' for the parameter ${\bf a}$ was postulated to be
\begin{eqnarray}
{
L\sub{ext}({\bf a}) = e^{-N \int_{Y} \overline{P}\sub{ev}({\bf y} | {\bf a}) \,d {\bf y}}
\prod_{k=1}^N \overline{P}\sub{ev}({\bf y}_k | {\bf a}),
\label{eq:extended_likeli}
}
\end{eqnarray}
where $N$ is the number of events in the data sample, and $\overline{P}\sub{ev}({\bf y} | {\bf a})$
stands for the probability to observe ${\bf y}$ without the requirement that
$\int_{Y} \overline{P}\sub{ev}({\bf y} | {\bf a}) \,d {\bf y}$ is normalized to 1.
Although this may not be immediately obvious, this formula actually presumes that
$\overline{P}\sub{ev}({\bf y} | {\bf a})$ contains an additional parameter:
a~freely floating normalization factor. Likelihood maximization with respect 
to that factor restores proper normalization of the $\overline{P}\sub{ev}({\bf y} | {\bf a})$,
as illustrated in~\cite{ref:estrada_thesis}.

The utility of this particular technique is not clear: the
normalization integral $\int_{Y} \overline{P}\sub{ev}({\bf y} | {\bf a}) \,d {\bf y}$
is needed anyway in order to calculate the likelihood, and the parameter estimation
requires an extra minimization step (typically, $-\ln L({\bf a})$ is minimized
in order to maximize the likelihood). There is, however, a more useful extended likelihood
formulation which can actually exploit the correlation,
if any, between the observed number of events and the probability density shape.
A~detailed description of this formulation
can be found in~\cite{ref:extended_likeli}.

A more substantial difference between Eq.~\ref{eq:prob_integral} and the
channel probability expressions
used, {\it e.g.,} 
in~\cite{ref:me1,ref:me2,ref:singlet1,ref:singlet2} consists in the explicit
inclusion of the phase space efficiency term $\epsilon_{i} ({\bf x}, {\bf a})$
in Eq.~\ref{eq:prob_integral}.
It is indeed possible to write
\begin{eqnarray}
{
P_{i}({\bf y} | {\bf a}) = \frac{\Omega({\bf y})}{\sigma_{i}({\bf a}) A_{i}({\bf a})}
\int_{X_i} W_{i}'({\bf y} | {\bf x}, {\bf a})\,
|M_{i} ({\bf x}, {\bf a})|^{2} \,T_{i}({\bf x}, {\bf a}) \,d {\bf x}
\label{eq:prob_integral2}
}
\end{eqnarray}
without this efficiency term, assuming transfer functions
$W_{i}'({\bf y} | {\bf x}, {\bf a})$ of a different kind. In order to maintain proper likelihood
normalization, these transfer functions must satisfy a different
normalization condition: $\int_{Y} W_{i}'({\bf y} | {\bf x}, {\bf a})\, d {\bf y} = 1$
for every $i$, ${\bf x}$, and ${\bf a}$. It is obvious that the
integrals in Eqs.~\ref{eq:prob_integral}
and~\ref{eq:prob_integral2} will result in the same likelihoods calculated
for an arbitrary data sample in case
$W_{i}'({\bf y} | {\bf x}, {\bf a}) = W_{i}({\bf y} | {\bf x}, {\bf a}) 
\,\epsilon_{i} ({\bf x}, {\bf a})$.
This relationship can be multiplied by $\Omega({\bf y})$
and integrated over~${\bf y}$. Together with the respective normalization
conditions for $W_i$ and $W_{i}'$, this leads to an intuitive
definition of phase space efficiency in terms of
$W_{i}'({\bf y} | {\bf x}, {\bf a})$ and $\Omega({\bf y})$:
\begin{eqnarray}
{
\epsilon_{i} ({\bf x}, {\bf a}) = 
\int_{Y} \Omega({\bf y}) W_{i}'({\bf y} | {\bf x}, {\bf a})\, d {\bf y}.
\label{eq:norm3}
}
\end{eqnarray}
There is, however, a crucial difference between functions
$W_{i}({\bf y} | {\bf x}, {\bf a})$ and $W_{i}'({\bf y} | {\bf x}, {\bf a})$:
in order to completely define $W_{i}({\bf y} | {\bf x}, {\bf a})$, one has
to know the function values only for those ${\bf y}$ for which $\Omega({\bf y}) = 1$.
On the other hand, the functions $W_{i}'({\bf y} | {\bf x}, {\bf a})$ must
be defined for {\it every {\bf y}} so that their normalization integrals
can be evaluated. As it will be discussed in the next section,
no reliable algorithm has been identified so far for constructing
such transfer functions 
--- they
suffer from {\it underspecification} in the $\Omega({\bf y}) = 0$ region.

For the remainder of this article, the following terminology will be employed.
Transfer functions which satisfy the normalization condition 
 $\int_{Y} W_{i}'({\bf y} | {\bf x}, {\bf a})\, d {\bf y} = 1$ will be
called ``type I''. Historically, these transfer functions were introduced first.
Functions normalized by 
$\int_{Y} \Omega({\bf y}) W_{i}({\bf y} | {\bf x}, {\bf a})\, d {\bf y} = 1$
will be designated ``type II''. Finally, the ``type III''
transfer functions will be formally defined via the type~II functions by
the expression $W_{i}''({\bf y} | {\bf x}, {\bf a}) \equiv W_{i}({\bf y} | {\bf x}, {\bf a}) 
\,\epsilon_{i} ({\bf x}, {\bf a})$. These functions are normalized by
\begin{eqnarray}
{
\int_{Y} \Omega({\bf y}) W_{i}''({\bf y} | {\bf x}, {\bf a})\, d {\bf y} 
= \epsilon_{i} ({\bf x}, {\bf a}).
\label{eq:normtf}
}
\end{eqnarray}
Although this equation looks almost identical to Eq.~\ref{eq:norm3},
its logic is inverted:
instead of using the transfer function to determine
the phase space efficiency, the efficiency is used to impose
the transfer function normalization.
Type III transfer functions defined in this manner
do not suffer from the type I underspecification
problem, and these are the correct functions to use instead of $W_{i}'$
inside the channel probability
integral represented by Eq.~\ref{eq:prob_integral2}.

\section{Transfer Function Modeling}
\label{sec:tf}

In order to obtain type II (or any other) detector transfer functions directly from data, 
one has to solve the inverse problem for
$W_{i}({\bf y} | {\bf x}, {\bf a})$. Unfortunately, this is feasible only when the
phase space $X_i$ is sufficiently restricted and the background
contamination is minimal. Such conditions can indeed be found
in test beams and in certain low-background processes
on $e^{+}e^{-}$ colliders, but $pp$ or $p\bar{p}$ collider environments
are not conductive to this approach. Therefore, transfer function
derivation procedures utilized so far in HEP ME analyses always
relied upon a reasonably well-tuned detector
simulation software package. For a given channel, such a package
starts with ${\bf x}$ and provides either a simulated ${\bf y}$ or an indication
that the event does not fall inside
the ${\bf y}$ region for which $\Omega({\bf y}) = 1$.

Computationally, it is not feasible to estimate the transfer function
by simulation in the process of evaluating Eq.~\ref{eq:prob_integral} integral. 
For modern HEP experiments, it takes a~few tens of CPU seconds to simulate
particle detector response to an
event\footnote{Typical for ATLAS and CMS detector simulation packages.}.
Since at least a~few hundred events are needed to
form an estimate of the probability density for each phase space point
and the number of phase space points per integral can be about 10$^5$, the required
CPU load is prohibitive. Instead, one constructs the transfer functions
from pools of pre-simulated Monte Carlo (MC) events by introducing reasonable
assumptions about the transfer function properties. The main assumptions
are:
\begin{thinlist}
\item To the first order, the full event transfer function
      can be factorized into the product of transfer functions
      for individual physics objects. Small corrections, {\it e.g.,}
      for nearby jets, can be introduced on top of this factorization.
\item The transfer functions are piecewise continuous together with their first
      few derivatives. The discontinuity points are known in advance
      (for example, at the junctions of separate detector subsystems).
\end{thinlist}
Usually, the detector response to jets is the most difficult transfer function
factor to model. A~significant fraction of ME analyses performed so far
attempted to construct type I jet transfer functions by following the ``original
recipe'' outlined in~\cite{ref:estrada_thesis}. According to this recipe,
jets are generated by MC with the same process as the process under study
(this makes sense for factorized transfer functions without additional
corrections, as energy
response for most jet reconstruction algorithms does exhibit
a mild dependence on jet multiplicity).
The selection of events is performed either according to $\Omega({\bf y})$
used for the data or according to a~different $\Omega'({\bf y})$ which
corresponds to modified, 
``relaxed''\footnote{$\Omega'({\bf y}) = 1$ for every {\bf y} for which
$\Omega({\bf y}) = 1$. In addition, $\Omega'({\bf y}) = 1$ for some
{\bf y} for which $\Omega({\bf y}) = 0$.} selection 
criteria. Simulated jets
are matched to Monte Carlo partons using an angular matching criterion
in the $\eta$-$\varphi$ space. The jet angular resolution is assumed to
be perfect (modeled by Dirac delta functions), while a~normalized probability density
function (a double Gaussian is common) is
fitted to a distribution of some energy-like {\it jet} quantity (transverse momentum or energy)
using the same (or sometimes different) energy-like {\it parton} quantity
as the explanatory variable (predictor). Some functional form
(usually linear) is assumed for the
parameters of the fitted density expressed in terms of the predictor.
This whole procedure thus fits a nonlinear regression model in which
the shape of the response distribution is constrained by its assumed functional form
and depends on the predictor.

As it turns out, this recipe has a serious
deficiency and it can't possibly produce
fully correct type~I transfer functions. The problem is that
the Monte Carlo events in the fits
are preselected, and only events with those ${\bf y}$ for which
$\Omega({\bf y}) = 1$ (or $\Omega'({\bf y}) = 1$, depending on the details)
are used. Thus the fitted MC events do not populate the complete $Y$ space
on which type I transfer function normalization is defined.
Such event samples can instead be used to fit type II transfer functions,
but the recipe calls for a wrong functional form which can't
represent a sharp cutoff in~${\bf y}$. Even if the
sample selection criteria are relaxed
as much as possible, the inherent inefficiencies of jet reconstruction
algorithms would introduce an effective selection $\Omega''({\bf y})$ 
which should not be ignored.

Defective transfer functions derived according
to this recipe affect probabilities of
events which contain low-energy jets near the ${\bf y}$ cutoff
but do not invalidate the overall measurement results.
The saving grace is provided by
the analysis final calibration procedure which corrects for biases and pulls.
However, just as
unnecessary dimensionality reduction assumptions, such functions degrade
either the measurement precision or the power of the signal discriminant.

Other methods of transfer function derivation have been developed.
In Ref.~\cite{ref:freeman} type I transfer functions were constructed
by an altered procedure. The deficiency of the original recipe
was noted and the $\Omega'({\bf y}) = 1$ region was used to fit
just the shapes of the transfer function densities but not to fix their
normalization. The transfer functions were then extrapolated
into the $\Omega'({\bf y}) = 0$ region devoid of events.
This method results in more appropriate functional shapes
and makes the transfer function normalizable
in the whole $Y$ space, but at the cost of introducing another problem: that of
extrapolation ambiguity. The fatter the extrapolated tail inside $\Omega'({\bf y}) = 0$, the
smaller the function values become inside $\Omega'({\bf y}) = 1$
due to the imposed normalization condition. For this method, the effect of
transfer function mismodeling is difficult to assess. Nevertheless,
this is undoubtedly an improvement upon the original recipe.

The authors of Ref.~\cite{ref:shief} explicitly build type II
transfer functions using appropriate functional shapes
which can model the $\Omega({\bf y})$ cutoff. However, these functions are then used
as if they were type~I, inside Eq.~\ref{eq:prob_integral2}. The
authors demonstrate that in this case the overall probability normalization
integral (called ``observable cross section'' in their paper) becomes independent of
detector acceptance. Instead of raising the red flag, this property
appeals to the authors because detector-related parameters
no longer enter the likelihood normalization. It should be obvious to the reader
of this note that this ``simplification'' of the likelihood 
comes at the price of an inappropriate transfer function model which distorts the
statistical description of the process and increases the measurement error.

A proper construction of type III jet transfer functions was
carried out in Ref.~\cite{ref:plujan} (at the time of this writing,
this work stands as the most precise single measurement of the top quark mass).
The authors model both transverse momentum and angular jet response with
nonparametric statistical techniques. They also employ a nonparametric estimate
of the phase space efficiency to normalize their transfer functions
in the $\Omega({\bf y}) = 1$ region. The resulting transfer functions
are then used to calculate event observation probabilities according to
Eq.~\ref{eq:prob_integral2}.

\section{Practical Consequences}

It should be apparent from the preceding discussion that an estimate
of the phase space efficiency $\epsilon_{i} ({\bf x}, {\bf a})$
must be constructed in order to reliably calculate the probability
$P_{i}({\bf y} | {\bf a})$ to find an event with observables ${\bf y}$. The calculation
can utilize either type~II transfer function in combination with
Eq.~\ref{eq:prob_integral} or type~III transfer function in combination  with
Eq.~\ref{eq:prob_integral2}.
Although these approaches are technically equivalent, the one
which utilizes Eq.~\ref{eq:prob_integral} is conceptually easier to understand
because in that equation $\epsilon_{i} ({\bf x}, {\bf a})$ has a very 
straightforward probabilistic interpretation.

The $\epsilon_{i} ({\bf x}, {\bf a})$ estimate
can be obtained from the usual Monte Carlo event
samples generated for analysis calibration purposes.\footnote{A number of statistical
techniques can be employed for this purpose. Local quadratic logistic regression 
is the personal favorite of the author.}
It is important that these event samples {\it are not filtered}, as
often done in order to conserve disk space.
This is because
reconstruction of the ${\bf x}$-dependent efficiency denominator
($\epsilon$ = events accepted/events total) becomes difficult or impossible
when a~fraction of events in the sample is discarded based on the value of~${\bf y}$.

Finally, it should be pointed out
that both type II and type III transfer functions
can be constructed using ``relaxed'' MC event selection criteria ({\it e.g.},
in order to simplify
normalization of the transfer functions when certain types of instrumental
parameters, such as the overall jet energy scale, are scanned).
In this case the corresponding ``relaxed'' phase space efficiencies should
be derived and used inside Eqs.~\ref{eq:prob_integral} or~\ref{eq:normtf}.

\end{document}